\def\av{$A_{\rm V}$}
\def\ha{H$\alpha$}
\def\wha{$W_{\rm H\alpha}$}

\def\st{[S\,{\sc ii}]}
\def\nt{[N\,{\sc ii}]}

\def\eb{$E_{\rm b}$}

\documentclass[iop,appendixfloats]{emulateapj}
\slugcomment{Accepted for publication in ApJL}

\shorttitle{The Dust Attenuation Law in Distant Galaxies}
\shortauthors{Kriek \& Conroy}

\begin{document}
  
\title{The Dust Attenuation Law in Distant Galaxies: Evidence for Variation with Spectral Type}

\author{Mariska Kriek\altaffilmark{1}, \& Charlie Conroy\altaffilmark{2}}

\altaffiltext{1}{Astronomy Department, University of California at Berkeley, Berkeley, CA 94720}

\altaffiltext{2}{Department of Astronomy \& Astrophysics, University of California at Santa Cruz, Santa Cruz, CA 95064}

\begin{abstract}
This letter utilizes composite spectral energy distributions (SEDs)
constructed from NEWFIRM Medium-Band Survey photometry to constrain
the dust attenuation curve in $0.5<z<2.0$ galaxies. Based on
similarities between the full SED shapes (0.3-8$\mu$m), we have
divided galaxies in 32 different spectral classes and stacked their
photometry. As each class contains galaxies over a range in redshift,
the resulting rest-frame SEDs are well-sampled in wavelength and show
various spectral features including H$\alpha$ and the UV dust bump at
2175\,\AA. We fit all composite SEDs with flexible stellar population
synthesis models, while exploring attenuation curves with varying
slopes and UV bump strengths. The Milky Way and Calzetti law provide
poor fits at UV wavelengths for nearly all SEDs. Consistent with
previous studies, we find that the best-fit attenuation law varies
with spectral type. There is a strong correlation between the best-fit
dust slope and UV bump strength, with steeper laws having stronger
bumps. Moreover, the attenuation curve correlates with specific
star formation rate (SFR), with more active galaxies having shallower
dust curves and weaker bumps. There is also a weak correlation with
inclination. The observed trends can be explained by differences in
the dust-to-star geometry, a varying grain size distribution, or a
combination of both. Our results have several implications for galaxy
evolution studies. First, the assumption of a universal dust model
leads to biases in derived galaxy properties. Second, the presence
of a dust bump may result in underestimated values for the UV slope,
used to correct SFRs of distant galaxies.
\end{abstract} 

\keywords{dust, extinction --- galaxies: stellar content}

\section{INTRODUCTION}\label{sec:int}

After dark matter, dust may be the most mysterious component of
galaxies. Aside from the Milky Way (MW) and a few well-studied
galaxies, we know little about the types of grains, the distribution
of dust with respect to stars, or the dust to gas or dust to star
ratio. Consequently, the dust production and destruction cycle and the
role of dust within the baryon cycle of galaxies are poorly
understood \citep[e.g.,][]{dr09,ma09,ga11}. We know that dust affects
the energy balance in the interstellar medium (ISM), the cooling of
gas, and hence star formation. However, dust is currently not included
in theoretical galaxy evolution models, and appropriate observations
needed to constrain these processes are still missing. Moreover, in
order to study the stellar populations of galaxies, we must have a
solid understanding of the properties and spatial distribution of
dust.

When deriving the properties of distant galaxies from photometric or
spectroscopic observations, we commonly assume a uniform attenuation
law, often implemented as a simple screen. Yet we know that the dust
attenuation law is not universal. For example, the MW and the Large
Magellanic Cloud (LMC) extinction curve to a lesser extent exhibit a
dust absorption feature around 2175\,\AA\ \citep{st65}, which is
absent in 4 out of 5 sight-lines toward the Small Magellanic Cloud
(SMC) and in an empirical attenuation curve for starburst
galaxies \citep{ca00}. This feature has been seen in other galaxies as
well, out to $z\sim1$ or even higher, though not as strong as for the
Milky Way
\citep[e.g.,][]{mo02,bu05,no09,el09,co10b,co10c,wi11,bu11}. The slope of
the attenuation curve also varies among different sight-lines and
galaxies.

Variations in the dust attenuation curve are expected on theoretical
grounds \citep{wg00}. However, the origin of the observed differences
between the various laws are poorly understood. For example, we do not
know what types of grains are responsible for the 2175\,\AA\ dust bump
\citep[e.g.,][]{dr93}, nor what physical process governs its
strength. We stress that dust attenuation and extinction are distinct
concepts; attenuation includes dust geometry and radiative transfer
effects, which may in part explain the observed differences.

\begin{figure*}  
  \begin{center}   \includegraphics[width=0.88\textwidth]{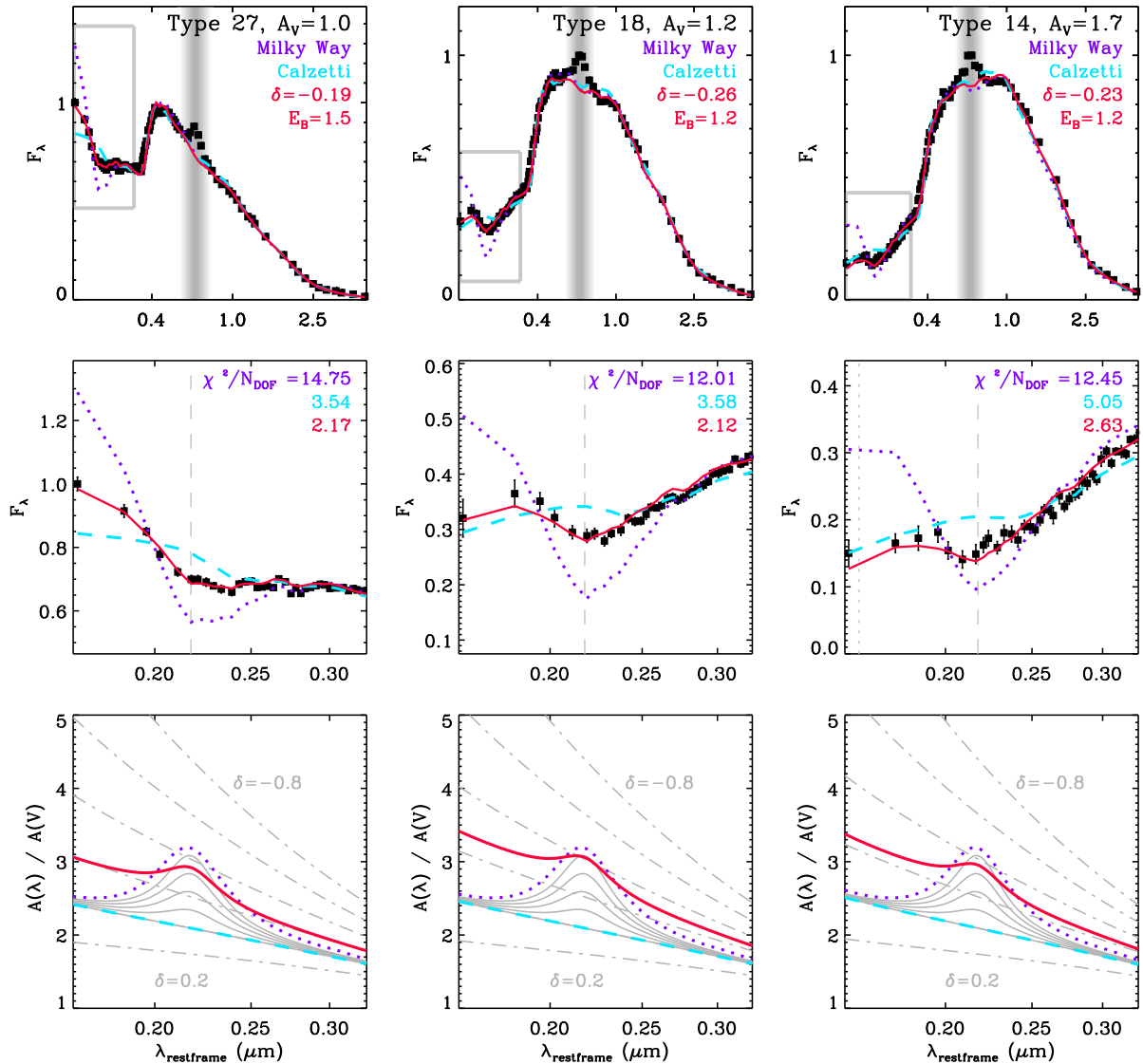} \caption{Composite
  SEDs for three arbitrarily chosen galaxy types, ordered by
  increasing value of $A_V$. The top panels show the full wavelength
  range and the middle panels zoom in on the gray open box in the
  rest-frame UV. The gray shaded area indicates the region
  around \ha. The best-fit FSPS models for different dust attenuation
  curves are overplotted. We explore the Calzetti law (light blue),
  the MW law (purple), and a dust law in which the slope ($\delta$)
  and strength of the UV bump ($E_b$) are free parameters (red). The
  corresponding attenuation curves are shown in the bottom panels. The
  parametrization of the slope ($\delta = [-0.8,0.2]$) and bump
  strength ($E_b=[0,4]$) are represented by the dashed-dotted and
  solid gray curves in the bottom panels. Both the Calzetti and the MW
  dust law provide poor fits in the rest-frame UV. \label{fig:fit}}

  \end{center}
\end{figure*}

In order to understand the role of dust and to correct the integrated
light from galaxies, we need to gain a better understanding of the
processes shaping the dust attenuation curve. The dust attenuation
curve can be derived from the stellar light of galaxies, but it
requires wide wavelength coverage and sufficient spectral
resolution. Both requirements are needed to break degeneracies with
other modeling parameters, such as stellar age and star formation
timescale \citep[e.g.,][]{la05} and to directly detect the UV dust
bump. This combination has recently become feasible with the
completion of the NEWFIRM Medium-Band Survey
\citep[NMBS;][]{wh11,vd09}. In this Letter we use well-sampled
composite SEDs constructed from the NMBS photometry \citep{kr11}, in
combination with the flexible stellar population synthesis
\citep[FSFS;][]{co09,co10a,cg10} models, to constrain the dust attenuation
curve in $0.5<z<2.0$ galaxies.

\section{DATA}

The NMBS is a 0.5 square degree photometric survey in the COSMOS and
AEGIS fields, unique for its 5 customly designed medium-band
near-infrared filters. The data obtained using these filters are combined
with optical-to-MIR photometry. For this study we make use of the NMBS
catalogs in the COSMOS field, which include medium-band photometry at
optical wavelengths as well. When no spectroscopic redshifts are
available, we use photometric redshifts \citep[EAzY;][]{br08}, which
have an accuracy of $\Delta z/ (1+z) <
0.02$ \citep{wh11,kr11,br11}. The high quality of the NMBS
photometry has been illustrated in \cite{vd10} and \cite{wh10,wh12}.

In \cite{kr11} we divided the NMBS galaxies at $0.5<z<2.0$ and with a
S/N$_{\rm K}> 25$ into different spectral classes, based on their
observed optical-to-IRAC SEDs. Each spectral class contains 22-455
individual galaxies of similar spectral shape. For each class we
constructed a composite SED, by de-redshifting and scaling the
observed photometry. These composite SEDs are basically low-resolution
spectra ($R\lesssim25$, Fig.~\ref{fig:fit}), and show many spectral features among
which \ha\ which is blended with [N\,{\sc ii}] and [S\,{\sc ii}] and
the blended [O\,{\sc iii}] and H$\beta$\ lines. Continuum and
absorption features are also visible, among which Mg\,{\sc ii}, the
Balmer and 4000\,\AA\ breaks, and the dust absorption feature at
2175\,\AA. 

\begin{figure*}  
  \begin{center}  
  \includegraphics[width=0.89\textwidth]{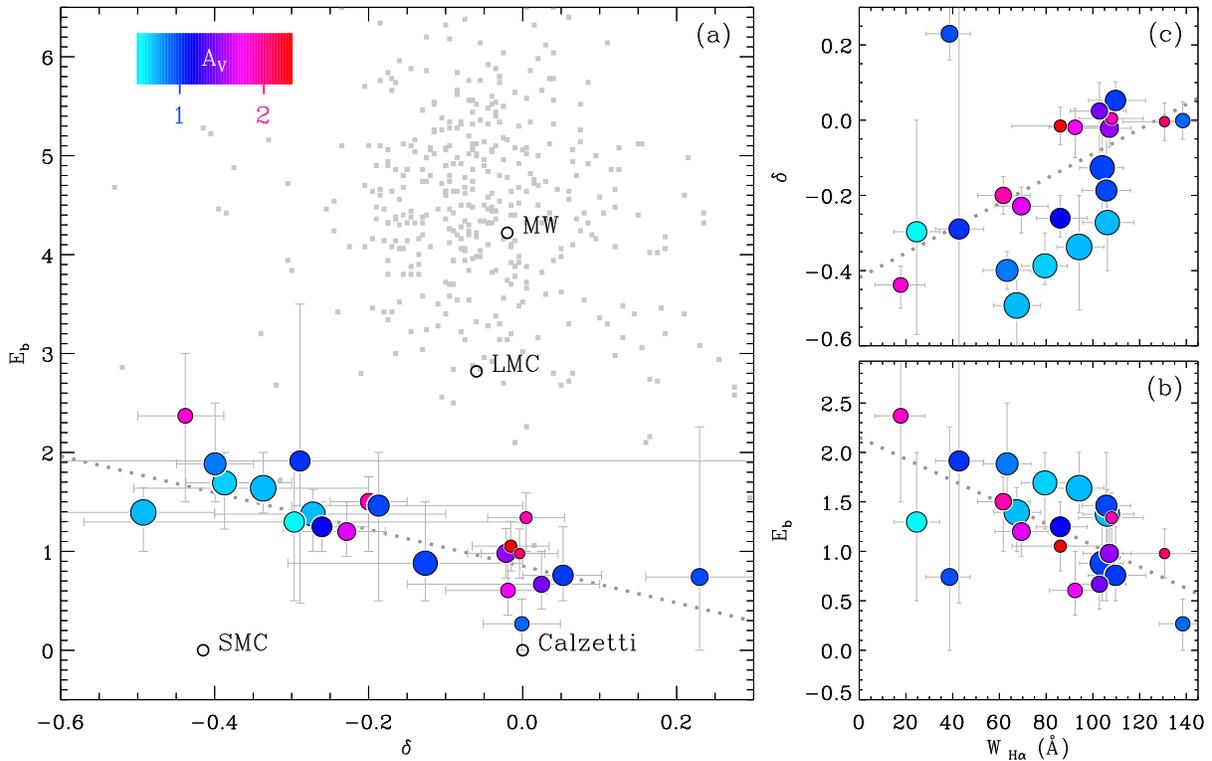}\\

  \caption{{\bf a)} The best-fit values for the bump strength ($E_B$)
    and the slope ($\delta$) of the dust attenuation curve for the
    different SED types. The symbol size reflects the number of
    galaxies used to construct the composite SED, and the color
    indicates the best-fit value for $A_V$ (given the best-fit dust
    law). Galaxy types with $A_V < 0.5$ (which have primarily
    quiescent stellar populations) are not included in this figure, as
    the dust law could not be constrained well. The Calzetti, MW, LMC
    and SMC attenuation and extinction curves are indicated by the
    open circles, and the small gray squares represent the MW
    sight-lines by \cite{va04}. There is a clear correlation between
    the slope of the attenuation curve and the UV bump strength. {\bf
    b-c)} The bump strength and slope versus the \ha\ equivalent
    width. The dust attenuation curve is shallower and the UV bump is
    weaker for galaxies with higher \wha\ (i.e., higher specific
    SFRs).\label{fig:results}}

  \end{center}
\end{figure*}

\section{CONSTRAINING THE DUST ATTENUATION LAW}

We constrain the dust attenuation curve by fitting the composite SEDs
with FSPS models in their default setting. These settings include a
low contribution from thermally-pulsing asymptotic giant branch
stars \citep[e.g.,][]{cg10,kr10,zi13}. We assume a \cite{ch03} initial
mass function (IMF) and an exponentially declining star forming
history. Our result is robust against the choice of the IMF, as it
primarily affects the mass-to-light ratio, and has little impact on
the shape of the SED. We explore a grid of stellar ages, star
formation timescales, and three different metallicities (0.0096, 0.019
($Z_{\odot}$), \& 0.03). We use the stellar population fitting code
FAST \citep{kr09}, which is based on a simple $\chi^2$
minimization. For the photometric uncertainties we assume a flat error
in $F_\lambda$ of 3\% of the average flux, in order to ensure that the
full SED shape will be taken into account in the fit. If we would use
the formal errors, the fit would be driven completely by the very
small error bars at longer wavelengths. The region around \ha\ is
masked in the fit. The confidence intervals are calibrated using Monte
Carlo simulations.

\begin{figure*}  
  \begin{center}  
  \includegraphics[width=0.99\textwidth]{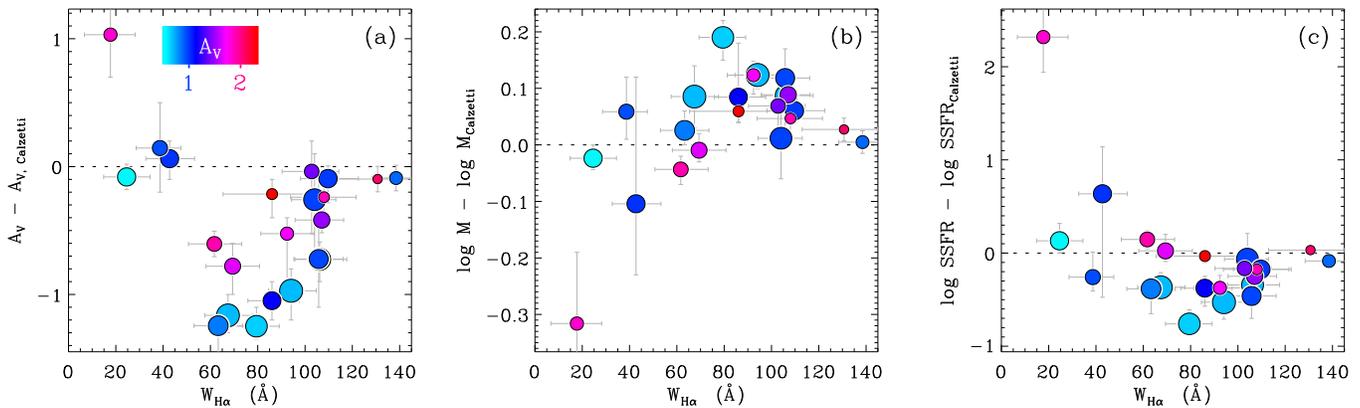}

  \caption{The difference in derived galaxy properties (\av, stellar
    mass, specific SFR) between a free dust attenuation curve and the
    Calzetti law. The symbols are similar as in
    Figure~\ref{fig:results}. These panels illustrate that a universal
    dust law, implemented as a simple screen, may result in large
    systematic biases in derived galaxy
    properties.\label{fig:implications}}

  \end{center}
\end{figure*}

First, we explore the MW \citep{ca89} and the Calzetti dust
attenuation curves, and find the best-fit stellar population model
when allowing all other parameters to vary. Next, we parameterize the
dust attenuation curve following the prescription by \cite{no09}:

\begin{equation}
A(\lambda) = \frac{A_V}{4.05} (k'(\lambda) + D(\lambda))  \left(\frac{\lambda}{\lambda_V}\right) ^\delta 
\label{eq:slope}
\end{equation}
with $D$ the Lorentzian-like Drude profile to parameterize the UV bump, defined as: 
\begin{equation}
D(\lambda) = \frac{E_b (\lambda\,\Delta \lambda)^2 }{(\lambda^2-\lambda_0^2)^2 + (\lambda\,\Delta \lambda)^2}
\label{eq:bump}
\end{equation}
For the central wavelength of the dust bump $\lambda_0$ we adopt
2175\,\AA. For $\Delta \lambda$, the full width at the half maximum of
the bump, we assume a value of 350\,\AA\ as found by
\cite{no09} from a stack of a spectra of high-redshift
galaxies and by \cite{se79} for the MW UV bump. We explore a grid of
values for \eb\ and $\delta$, while leaving all other fitting
parameters free. All attenuation curves are implemented as uniform
screens.

In Figure~\ref{fig:fit} we show the best-fit stellar population models
for three composite SEDs, for the three different attenuation
laws. This figure illustrates that for all three SEDs the Calzetti and
MW law provide poor fits at rest-frame UV wavelengths. When
leaving both the slope and bump strength as free parameters, we can
reproduce their full SED shapes.

The best-fit values for \eb\ and $\delta$ for all composite SEDs with
$A_V > 0.5$ are shown in Figure~\ref{fig:results}a. There is a clear correlation between both
properties, described by the following linear fit:
\begin{equation}
E_b = (0.85 \pm 0.09) - (1.9 \pm 0.4 ) \, \delta
\label{eq:corr1}
\end{equation}
SED types with steeper attenuation curves have stronger UV bumps, while
shallower attenuation curves go together with weaker UV bumps.

\section{CORRELATIONS WITH SPECTRAL TYPE}

In the previous section we found a large variety in dust attenuation
curves among the different SED types. In order to examine whether the
shape of the dust law correlates with spectral type, we show both
$E_{\rm b}$ and $\delta$ as a function of the equivalent width of \ha\
(\wha) in Figures~\ref{fig:results}b~and~c. Similar to the specific
SFR, \wha\ is a measure of the present to past star formation in a
galaxy. This measurement directly follows out of the composite SEDs,
and thus is independent of the FSPS modeling. We do note that \wha\
may include some contamination by \nt\ and \st, and thus may be
slightly overestimated.

Figures~\ref{fig:results}b~and~c show that the shape of attenuation
curve correlates with spectral type. Galaxies with higher values
for \wha\ have shallower attenuation curves with weaker UV bumps. The
only significant outlier to this trend is SED type 8, which is best
fit by a shallow dust slope ($\delta\sim0.2$), while having a
low \wha. The least square fits give the following relations:
\begin{equation}
E_b = (2.2 \pm 0.3) - (1.1 \pm 0.3 ) \, \left( \frac{W_{\rm
H\alpha}}{100 \rm \AA}\right)
\label{eq:corr2}
\end{equation}
\begin{equation}
\delta = (0.33 \pm 0.04 ) \, \left( \frac{W_{\rm H\alpha}}{100 \rm \AA}\right) - (0.41 \pm 0.04)
\label{eq:corr3}
\end{equation}

Our findings suggest that the dust attenuation curve should not be
treated as a universal function, but should be related to the spectral
type. In the next section we examine the implications of our results.

\section{IMPLICATIONS}

Comparing broadband SEDs or spectra with stellar population synthesis
(SPS) models is a popular method to derive galaxy properties. In this
procedure, the shape of the dust attenuation curve is almost always
held fixed. Here we examine possible biases introduced by assuming the
popular Calzetti dust attenuation law. Figure~\ref{fig:implications}
shows the difference in best-fit galaxy properties for a free dust law
compared to when assuming the Calzetti law. This figure
illustrates that a fixed dust law may introduce systematic biases.
\av\ and the specific SFR could be significantly
overestimated for galaxies with high \wha, while stellar mass may be
slightly underestimated. For lower values of \wha\ we find the
opposite effect. However, this trend is driven by only one SED type
(12), which was previously noticed in \cite{kr11} as being poorly fit by
any SPS model, and thus this trend is not significant.

\begin{figure}  
  \begin{center}  
  \includegraphics[width=0.45\textwidth]{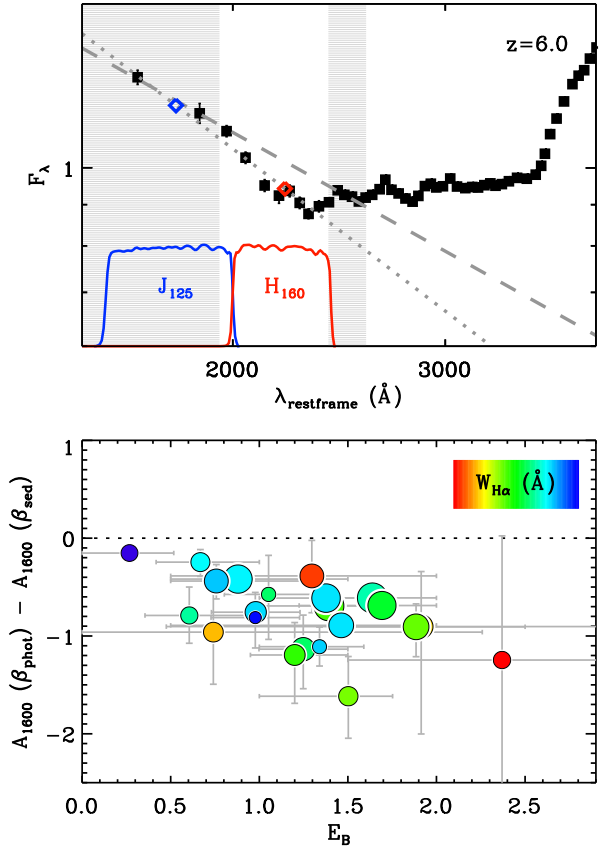}

  \caption{Implication for the dust correction based on the UV slope.
    The gray dashed line in the top panel represents the best-fit UV
    slope to the composite SED while masking the dust absorption
    feature. Only datapoints in the gray shaded areas are included in the
    fit, following the prescription by \cite{me99}. The gray dotted
    line shows the best-fit UV slope if we only had two photometric
    bands. In the lower panel we show the
    difference in derived values for A$_{1600}$ using the relation
    by \cite{me99} versus the strength of the dust feature for all
    composite SEDs with $A_V>0.5$. This figure illustrates that dust
    corrections based on photometric measurements, may be significantly
    underestimated, resulting in underestimated SFRs. However,
    the choice of filters in this illustration may possibly represent the
    worst case scenario, and different filters, redshifts, or the
    availability of multiple filters will lead to smaller
    biases.\label{fig:beta} }

  \end{center}
\end{figure}

At high redshift dust attenuation corrections are frequently derived
from the UV $\beta$ slope ($F_{\lambda}\propto\lambda^\beta$),
adopting the relation of \cite{me99}.  This relation was based on a UV
slope measured by masking the UV bump and other features.  In
contrast, the measurements at high redshift are based on a small
number of broadband filters \citep[e.g.,][]{re10,bo12}. In
Figure~\ref{fig:beta} we use the composite SEDs to assess how the UV
bump may influence the measurement of the UV slope, and hence the dust
correction. In the top panel we show an example composite SED, the
corresponding photometric measurements for two rest-frame UV filters
for this SED, and the slopes based on the composite SED (dashed line)
and photometric datapoints (dotted lines). In the lower panel we show
the difference in the derived attenuation correction -- based on the
two UV slopes -- versus the strength of the dust feature, for all
composite SEDs with \av$>0.5$. This figure illustrates that the
presence of a dust feature may lead to underestimated dust
corrections, and thus underestimated SFRs. Stronger dust features
clearly result in larger biases.

However, the illustration in Figure~\ref{fig:beta} is probably a worst
case scenario, where only two filters are used, and the reddest filter
completely overlaps with the UV bump. Several studies use three to
four photometric bands \citep[e.g.,][]{bo12} to measure the UV slope,
resulting in more robust measurements. Furthermore, \wha\ increases
with increasing redshift \cite[e.g.,][]{st13}, and thus at a redshift
of $z=6$ (used in the illustration in Fig.~\ref{fig:beta}) or higher
the bump may be insignificant. We do note though, that the utility of
the UV slope for estimating dust attenuation has been called into
question before \citep[e.g.,][]{ko04,jo07,co10b,go13}.

\begin{figure}  
  \begin{center}  
    \includegraphics[width=0.48\textwidth]{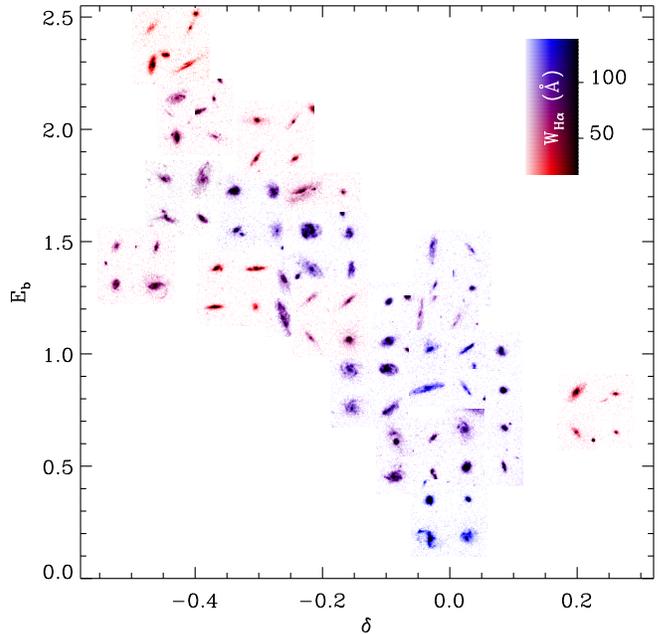} 

    \caption{The bump strength ($E_B$) vs. the slope ($\delta$) of the
      best-fit attenuation curve for all composite SEDs with
      $A_V>0.5$. Instead of symbols, we show the $i$-band HST images
      of 2-4 galaxies per SED type. The galaxies are chosen to be
      closest in redshift to $z=0.9$, thus the images show the
      rest-frame optical morphologies. The images are color coded by
      \wha\ (i.e, the specific SFR). This
      figure illustrates that galaxies with low \wha\ and/or edge-on
      galaxies have steeper attenuation curves with stronger UV bumps,
      while face-on and/or high \wha\ galaxies have shallower curves
      with weaker UV bumps.  \label{fig:im} }

  \end{center}
\end{figure}

\section{DISCUSSION}

The primary result of this Letter is that the shape of the dust
attenuation curve varies with spectral type. This trend may be
explained by differences in the dust-to-star geometry, which could
result in age-dependent extinction and variations in the scattering
and absorption of radiation \citep[e.g.,][]{ch13}. In particular, a
two-component dust model
\citep[e.g.,][]{ca94,cf00,gr00,wi11}, in which all stars are
attenuated by the diffuse ISM, while younger stars experience
additional attenuation due to dust in their short-lived birth clouds,
naturally produces the observed trends. This model predicts variation
with galaxy inclination as well, as the optical depth of the
diffuse component is larger for edge-on galaxies. In
Figure~\ref{fig:im}, we show HST morphologies \citep{sc07} as function
of $E_{\rm b}$ and $\delta$. This figure illustrates that face-on
and/or more active galaxies on average have shallower attenuation
curves with weaker UV bumps. Similar trends are also seen in local
galaxies \citep{wi11}. Only SED type 8, which we mentioned before,
does not follow the trends.

The observed correlations between the attenuation curve and the
specific SFR may also reflect variations in the grain size
distribution. This could arise from a changing balance between grain
formation, growth, and destruction with star formation
history \citep[e.g.,][]{go03}. See \cite{co13} for more discussion on
the proposed explanations

Our values for $E_b$ and $\delta$ are in the same range to those found
by \cite{bu12}, who fit the photometry of individual galaxies in the
GOODS-S field. They also find weaker UV bump strengths in galaxies
with larger specific SFRs \citep[see also][]{wi11}. \cite{bu12} find
no correlation though between the slope and the dust bump for their
sample. This may not be surprising, as for individual galaxies the
photometric resolution and S/N may not be sufficient to resolve the
bump and to break modeling degeneracies. As a result the scatter in
the measurements will be large, and possible trends may be washed out.

While the study presented in this Letter has been enabled by stacking
SEDs of similar type, this is also a weakness, as we may be combining
galaxies with different attenuation curves. Furthermore, the
rest-frame wavelength coverage is not the same for all redshifts, and
at bluer wavelengths the SEDs are dominated by the higher redshift
galaxies in our sample. Thus, extending this study to shorter
wavelengths, using GALEX data will give a more robust and
representative measurement of the UV attenuation curve.  Another
drawback is that we only look at the attenuated and not at the
re-emitted light. In a future study we will extend the composite SEDs
to longer wavelengths as well, so that we can simultaneously study the
dust attenuation and emission as a function of spectral type.

\section{SUMMARY}
  
In this Letter, we use composite SEDs, constructed from NMBS
photometry, to study variations in the dust attenuation curve in
$0.5<z<2.0$ galaxies. Based on similarities between the full SED
shapes, we have divided galaxies in different spectral classes and
stacked the photometry. As each spectral type contains galaxies which
span a range in redshift, the resulting SEDs are well sampled in
wavelength and show various spectral features among which H$\alpha$
and the 2175\,\AA\ dust feature. 

We fit all composite SEDs with FSPS models, while exploring attenuation
curves with varying slopes and UV bump strengths. The MW and
Calzetti law provide poor fits at UV wavelengths for nearly all SED
types. We find that the strength of the UV bump and the slope of the
curve are strongly correlated, with steeper slopes having stronger UV
dust bumps. Moreover, the shape of the dust attenuation curve
correlates with the equivalent width of \ha\ (i.e., specific SFR),
with higher \wha\ galaxies having shallower dust attenuation curves
and weaker UV bumps. The average attenuation law has a UV bump
strength of $\sim25$\% of the MW bump, and a slope in between
the SMC and the MW extinction curve.

Our results are consistent with a two-component dust model, in which
all stars are attenuated by the diffuse ISM, while young stars
experience extra attenuation due to dust in their short-lived birth
clouds. A qualitative analysis of the morphologies further supports
this picture, as edge-on galaxies have on average steeper dust curves
and stronger UV bumps. A varying grain size distribution may also
contribute to the observed correlations with H$\alpha$ equivalent
width.

A non-universal dust attenuation curve has numerous implications for
galaxy evolution studies. By assuming a universal dust law, derived
stellar population properties will suffer from systematic biases. For
example, for the Calzetti law, the dust content and specific SFR could
be significantly overestimated. Stellar mass measurements are more
robust, and are only slightly underestimated. Furthermore, the biases
vary with spectral type. The presence of the dust bump also
complicates dust corrections based the UV $\beta$ slope. In the worst
case scenario, when the reddest filter overlaps with the bump,
A$_{1600}$ could be underestimated by $\sim$1 mag, resulting in
significantly underestimated SFRs.  On a positive note, the observed
variation should motivate theoretical work aimed at embedding
self-consistent dust models into galaxy formation models.

\acknowledgements We thank the members of the NMBS and COSMOS teams
for releasing high-quality multi-wavelength data sets to the
community, and the referee for a constructive report. MK acknowledges
support from HST grant HST-AR-12847.01-A.  CC acknowledges support
from the Alfred P. Sloan Foundation.

\end{document}